\documentclass[
    reprint,
    showkeys,
    showpacs,
    amsmath,
    amssymb,
    aps,
    prb,
    twocolumn,
    floatfix
]{revtex4-2}
\usepackage{amsmath}
\usepackage{graphicx}
\usepackage{dcolumn}
\usepackage{bm}
\usepackage{braket}
\usepackage[caption=false]{subfig}
\usepackage{float}
\usepackage{booktabs}
\usepackage{multirow}
\usepackage{tabularx}
\usepackage{color}
\definecolor{blue}{RGB}{0,0,0}
\definecolor{link}{RGB}{57,106,177}
\usepackage{xr}
\newcommand{\ped}[1]{\ensuremath{_{\rm #1}}}
\newcommand{\apex}[1]{\ensuremath{^{\rm #1}}}

\usepackage{hyperref}
\hypersetup{
    colorlinks=true,
    linkcolor=link,
    citecolor=link,
    urlcolor=link
}
\AtBeginDocument{
\heavyrulewidth=.08em
\lightrulewidth=.05em
\cmidrulewidth=.03em
\belowrulesep=.65ex
\belowbottomsep=0pt
\aboverulesep=.4ex
\abovetopsep=0pt
\cmidrulesep=\doublerulesep
\cmidrulekern=.5em
\defaultaddspace=.5em
}
\begin{document}
\title{Ionic gating in metallic superconductors: A brief review}
\author{Erik Piatti}
\email{erik.piatti@polito.it}
\affiliation{\mbox{
Department of Applied Science and Technology, Politecnico di Torino, I-10129 Torino, Italy}
}


\begin{abstract}
Ionic gating is a very popular tool to investigate and control the {\color{blue}electric charge transport} and electronic ground state in a wide variety of different materials. This is due to its capability to induce large modulations of the surface charge density by means of the electric-double-layer field-effect transistor (EDL-FET) architecture, and has been proven to be capable of tuning even the properties of metallic systems. In this short review, I summarize the main results which have been achieved so far in controlling the superconducting (SC) properties of thin films of conventional metallic superconductors by means of the ionic gating technique. I discuss how the gate-induced charge doping, despite being confined to a thin surface layer by electrostatic screening, results in a long-range ``bulk" modulation of the SC properties by the coherent nature of the SC condensate, as evidenced by the observation of suppressions in the critical temperature of films much thicker than the electrostatic screening length, and by the pronounced thickness-dependence of their magnitude. I review how this behavior can be modelled in terms of proximity effect between the charge-doped surface layer and the unperturbed bulk with different degrees of approximation, and how first-principles calculations have been employed to determine the origin of an anomalous increase in the electrostatic screening length at ultrahigh electric fields, thus fully confirming the validity of the proximity effect model. Finally, I discuss a general framework -- based on the combination of \textit{ab-initio} Density Functional Theory and the Migdal-Eliashberg theory of superconductivity -- by which the properties of any gated thin film of a conventional metallic superconductor can be determined purely from first principles.\\\\
Cite this article as: E. Piatti \textit{Nano Ex.} \href{https://doi.org/10.1088/2632-959X/ac011d}{\textbf{2}, 024003} (2021).

\end{abstract}

\maketitle


\section{Introduction}
The electric field effect is the modulation of the electronic properties of a 
material by the application of a transverse electric field. Its most influential 
application is the field-effect transistor (FET), which forms the backbone of modern 
semiconductor electronics. The possibility to control the superconducting (SC) state 
by means of the electric field effect has been an open issue in condensed matter 
physics since the seminal experiments of Glover and Sherrill~\cite{GloverPRL1960} and 
Stadler~\cite{StadlerPRL1965} in the Sixties. In the former, one electrode of a capacitor was 
replaced with either a tin or an indium SC film, and upon charging the transition 
temperature $T\ped{c}$ of the films was shifted at most by $0.08$ and $0.07$~mK 
respectively, corresponding to a faint $\approx0.002\%$ shift in the original $T\ped{c}$~\cite{GloverPRL1960}. In the latter, the maximum $T\ped{c}$ shift was enhanced to $\approx 0.035\%$ by substituting the solid-oxide dielectric with 
a ferroelectric in the asymmetric capacitor~\cite{StadlerPRL1965}. These first results showed that controlling the SC phase of metallic superconductors via the electric field effect was feasible but of negligible impact for device applications, and the topic was largely abandoned for over thirty years. The discovery of high-$T\ped{c}$ superconductors led to a rekindled interest in the topic, since the much smaller carrier density of the cuprates with respect to metallic superconductors allowed a much more effective tunability. Examples included an electric-field-induced 10~K $T\ped{c}$ shift in YBa\ped{2}Cu\ped{3}O\ped{7-\delta} by gating through ultra-high-$\kappa$ dielectric SrTiO\ped{3}~\cite{MannhartAPL1993}, and a full superconductor-insulator transition (SIT) induced in GaBa\ped{2}Cu\ped{3}O\ped{7-\delta}~\cite{AhnScience1999} by ferroelectric gating. Similarly, the low carrier density of ultrathin Bi films was also exploited to achieve a SIT by gating through SrTiO\ped{3}~\cite{ParendoPRL2005}. Still, the range of applicability remained limited.

A turning point was achieved in the late 2000s with the introduction of the ionic 
gating technique, which rapidly became a very popular tool to investigate and control 
the electric transport and electronic ground state in a wide variety of different 
materials~\cite{YeNatMater2010, YeScience2012, JoNanoLett2015, ShiSciRep2015, YuNatNano2015, SaitoACSNano2015, GonnelliSciRep2015, LiNature2016, WangNature2016, OvchinnikovNatCommun2016, PiattiApSuSc2017, PiattiAPL2017, Gonnelli2dMater2017, ZengNanoLett2018, PiattiNL2018, DengNature2018, PiattiApSuSc2018mos2, WangNatNano2018, PiattiEPJ2019, RenNL2019, PiattiLTP2019, PiattiApSuSc2020}. Ionic gating operates by exploiting the ultrahigh electric fields that 
build up at a voltage-polarized solid-electrolyte interface in the so-called electric 
double layer (EDL), made possible by the sub-nanometric spacing between the two 
charged layers in the EDL~\cite{UenoJPSJ2014, FujimotoPCCP2013}. 
When incorporated in an EDL-FET architecture, this allows attaining modulations of 
the surface charge density in the device channel which are often comparable to those 
occurring in metallic systems~\cite{LiNature2016, DagheroPRL2012, TortelloApSuSc2013}. Since then, the ionic gating 
technique has found large success in tuning the properties of low- and 
moderate-carrier density superconductors. The SC properties of 
intrinsically-superconducting transition metal dichalcogenide NbSe\ped{2} have been 
efficiently tuned by ionic gating both in the electrostatic~\cite{XiPRL2016} and in 
the electrochemical~\cite{LiSUST2014, YoshidaAPL2016} regime. High-$T\ped{c}$ copper 
oxides are also strongly tunable by ionic gating, as showcased first in 
La\ped{1-x}Sr\ped{x}CuO\ped{4}~\cite{BollingerNature2011} and then in 
YBa\ped{2}Cu\ped{3}O\ped{7-\delta}~\cite{LengPRL2011, LengPRL2012}, although in 
oxides in general care has to be taken to distinguish between electrostatic 
field-effect doping and field-induced changes in the oxygen stoichiometry due to the 
high mobility of the oxygen vacancies~\cite{JeongScience2013, SchladtACSNano2013, WalterACSNano2016, ZhangACSNAno2017, ZengPRL2018, SaleemAMI2019}. Among iron-based 
superconductors, low-carrier-density compound FeSe has been shown to develop a 
high-temperature ($T\ped{c}\gtrsim 40$~K) SC phase upon surface electron 
accumulation~\cite{ShiogaiNatPhys2016, LeiPRL2016, HanzawaPNAS2016, MiyakawaPRM2018}, 
and bulk gate-induced lithiation~\cite{LeiPRB2017} and protonation~\cite{CuiCPL2019}. 
While these results have been shown to be somewhat transferable to 
FeSe\ped{1-x}Te\ped{x}~\cite{KounoSciRep2018, ZhuPRB2017} and 
(Li,Fe)OHFeSe~\cite{LeiPRB2016}, other iron-based compounds such as 
BaFe\ped{2}As\ped{2} have been shown to be much less compliant~\cite{PiattiPRM2019}, 
indicating that the ionic-gate tunability of iron-based superconductors is strongly 
dependent on the specific properties of each compound. 

In contrast, the applicability of the ionic gating technique to conventional metallic superconductors has received significantly less attention, likely owing to their high intrinsic carrier density preventing the observation of large gate-induced changes to the intrinsic $T\ped{c}$ or other strong changes to the ground state of the gated material. In this short review, I present the main results obtained so far concerning the tuning of the SC properties of metallic superconductors from both the experimental and the theoretical points of view, showing how the comparable simplicity of conventional metallic superconductors with respect to cuprates or iron-based compounds makes them an ideal playground to investigate the general physics of gate-induced perturbations to the SC state.

\section{Ionic-gate-tuned superconductivity in niobium and niobium nitride films}

\begin{figure*}
\centering
\includegraphics[width=0.8\textwidth]{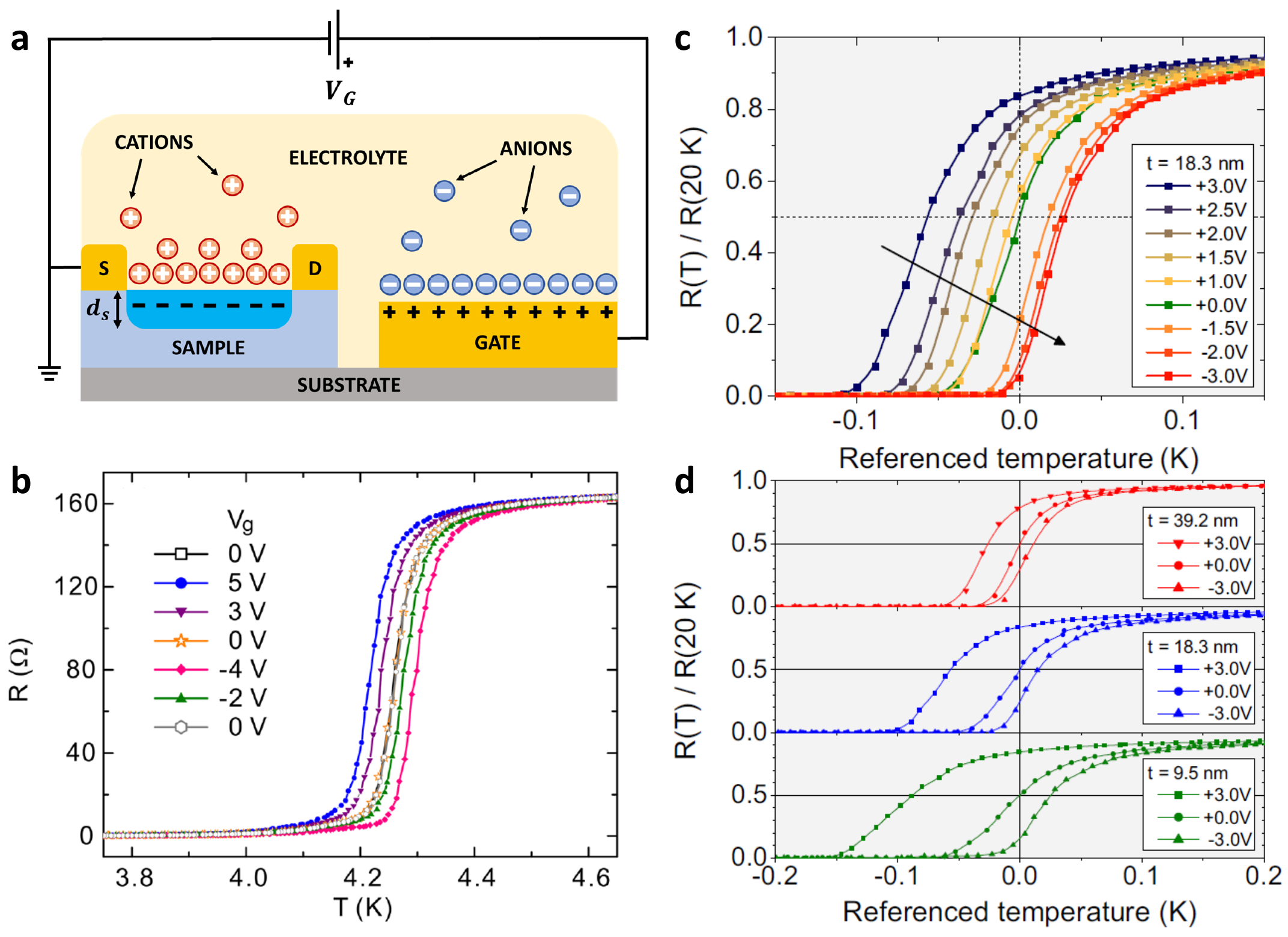}
\caption{
	\textbf{a}, Sketch of an EDL-FET with a metallic active channel, biased with a positive gate voltage $V\ped{G}$. Gate, source (S) and drain (D) electrodes are indicated, as well as the thickness of the surface charge-doped layer $d\ped{s}$.
	\textbf{b}, Resistance vs. temperature in the vicinity of $T\ped{c}$ of a 8nm-thick Nb thin film for different values of $V\ped{G}$. Reprinted from Ref.~\onlinecite{ChoiAPL2014}, with the permission of AIP Publishing.
	\textbf{c}, Normalized resistance vs. temperature in the vicinity of $T\ped{c}$ of a 18.3nm-thick NbN film for different values of $V\ped{G}$.
	\textbf{d}, Normalized resistance vs. temperature in the vicinity of $T\ped{c}$ of a NbN film for the extremal values of $V\ped{G}$, as a function of the total film thickness.
	In both \textbf{c} and \textbf{d} the temperature scale is referenced to the midpoint of the transition at $V\ped{G}=0$ (see Ref.~\cite{PiattiPRB2017} for details). \textbf{c},\textbf{d} are reprinted with permission from Ref.~\onlinecite{PiattiPRB2017}. Copyright 2017 by the American Physical Society.
}
\label{fig:SC_tuning}
\end{figure*}

The working principle of the EDL-FET architecture for an ion-gated metallic superconductor is sketched in Fig.~\ref{fig:SC_tuning}a. The active channel is separated from the gate counter-electrode by an electrolyte, which is usually either an ionic liquid~\cite{ChoiAPL2014} or an ion gel~\cite{PiattiJSNM2016, PiattiPRB2017}. When a positive/negative voltage $V\ped{G}$ is applied to the gate electrode, cation/anions in the electrolyte accumulate at the surface of the sample and lead to the accumulation/depletion of electrons at its surface to maintain charge neutrality, building up the EDL. If no bulk electrochemical interactions are triggered (such as intercalation, protonation or formation of mobile vacancies), the application of an ionic gate thus results in a strong charge doping which is confined to a thin layer from the surface, with a thickness $d\ped{s}$ that is set by the electrostatic screening length of the system, while the remaining bulk remains unperturbed~\cite{ChazalvielBook}. The resulting modification to the electric transport properties of both the normal and SC state in the active channel can then be monitored by standard four-point resistivity measurements~\cite{ChoiAPL2014, PiattiJSNM2016, PiattiPRB2017}.

The first metallic superconductor to be investigated by means of the ionic gating technique was niobium (Nb), a standard electron-phonon superconductor with a bulk $T\ped{c}=9.26$~K. Specifically, Choi et al. reported that the $T\ped{c}$ of niobium thin films could be reversibly modulated by gating through an ionic liquid~\cite{ChoiAPL2014}. The corresponding resistive transitions for an 8nm-thick film are shown in Fig.~\ref{fig:SC_tuning}b. From a qualitative standpoint, the results mirrored the early findings in indium thin films~\cite{GloverPRL1960}, with the accumulation of extra electrons in the system reducing the $T\ped{c}$ and the depletion of electrons enhancing it. The magnitude of the effect was of course much larger thanks to the huge capacitance of the ionic gate, with the largest $T\ped{c}$ shifts being around $\sim35$~mK, i.e. $\approx 0.8\%$ of the intrinsic $T\ped{c}\simeq4.3$~K~\cite{ChoiAPL2014}.
The shape of the resistive transition was unaffected by the applied gate voltage, indicating that the ionic gate did not introduce additional inhomogeneity in the sample. Notably, the gate-induced modulations of $T\ped{c}$ were volatile, i.e. fully reversible by simply removing the applied gate voltage and warming the sample above the melting point of the ionic liquid~\cite{ChoiAPL2014}. Additionally, it was also shown that the critical current $I\ped{c}$ was also tunable by the ionic gate with the same behavior of $T\ped{c}$, indicating that not only the onset of the SC state, but also the superfluid density, could be enhanced and suppressed by electron depletion and accumulation respectively~\cite{ChoiAPL2014}.

A second metallic superconductor, niobium nitride (NbN), was investigated via ionic gating shortly afterwards~\cite{PiattiJSNM2016, PiattiPRB2017}. With respect to pure Nb, NbN features a higher bulk $T\ped{c}\simeq 16$~K (making it more appealing for applications such as bolometers and single-photon detectors) and a shorter coherence length that allows thinning NbN films to few nanometers in thickness without major losses in $T\ped{c}$, all the while being mechanically robust and chemically stable~\cite{ChockalingamPRB2008}. Furthermore, preliminary \textit{ab initio} calculations suggested that its $T\ped{c}$ could be nearly doubled if a sufficiently large number of electrons were to be added to its first unit cell from the surface~\cite{PiattiJSNM2016}. In contrast with these predictions, it was experimentally shown that the gate-induced modulations of $T\ped{c}$ in NbN thin films followed the same qualitative behavior observed in Nb thin films: As shown in Fig.~\ref{fig:SC_tuning}c, electron depletion and accumulation enhanced and suppressed $T\ped{c}$ respectively, without affecting the shape of the transition, and the gate-induced shifts were fully volatile~\cite{PiattiJSNM2016, PiattiPRB2017}. From a quantitative standpoint, a larger maximum $T\ped{c}$ shift $\sim 85$~mK was reported in 9.5nm-thick NbN films, which however corresponded to a smaller $\approx 0.6\%$ relative modulation due to the larger intrinsic $T\ped{c}\simeq 13.7$~K with respect to that of Nb~\cite{PiattiPRB2017}.

In both ion-gated Nb and NbN thin films, the rigid shift of the resistive transition and the volatility of the $T\ped{c}$ modulation were strongly suggestive of a (mostly) electrostatic origin of the charge doping. On the other hand, the observation of $T\ped{c}$ suppressions in films of both materials with a thickness much larger than the electrostatic screening length is hardly compatible with a perturbation of the SC state occurring only at the surface, since the underlying unperturbed bulk would always reach a zero-resistance state at the same original $T\ped{c}$. Indeed, for $T\ped{c}$ suppressions to be detected in the resistive transition, a suppression of the SC pairing must extend across the whole thickness of the film. In Ref.~\onlinecite{ChoiAPL2014}, the observation of $T\ped{c}$ suppressions in NbN films as thick as 120 nm could not be justified in terms of pure surface charge doping, and the authors suggested that electro-mechanical effects due to the frozen ionic liquid may be responsible for the bulk modulations of SC. Conversely, corroborated by the fact that analogous bulk-like suppressions of $T\ped{c}$ had been reported in gated indium films in the absence of any measurable strain~\cite{GloverPRL1960}, in Ref.~\onlinecite{PiattiJSNM2016} it was suggested that the electrostatic screening responsible for the surface confinement of the charge doping may deviate strongly in a superconductor from that found in a normal metal. Additionally, in Ref.~\onlinecite{PiattiPRB2017} it was shown that the gate-tunability of $T\ped{c}$ was strongly enhanced upon reducing of the thickness of the NbN thin films by means of ion milling (Fig.~\ref{fig:SC_tuning}c). This ruled out a pure bulk origin of the modulation, which would be thickness-independent in the case of $T\ped{c}$ suppressions, as well as a purely surface-bound phenomenon, which would be thickness-independent in the case of $T\ped{c}$ enhancements, and hinted at a more complex interplay between surface charge doping and bulk perturbation of the SC pairing~\cite{PiattiPRB2017}.

\section{Surface-to-bulk coupling via superconducting proximity effect}
\begin{figure}
\centering
\includegraphics[width=\columnwidth]{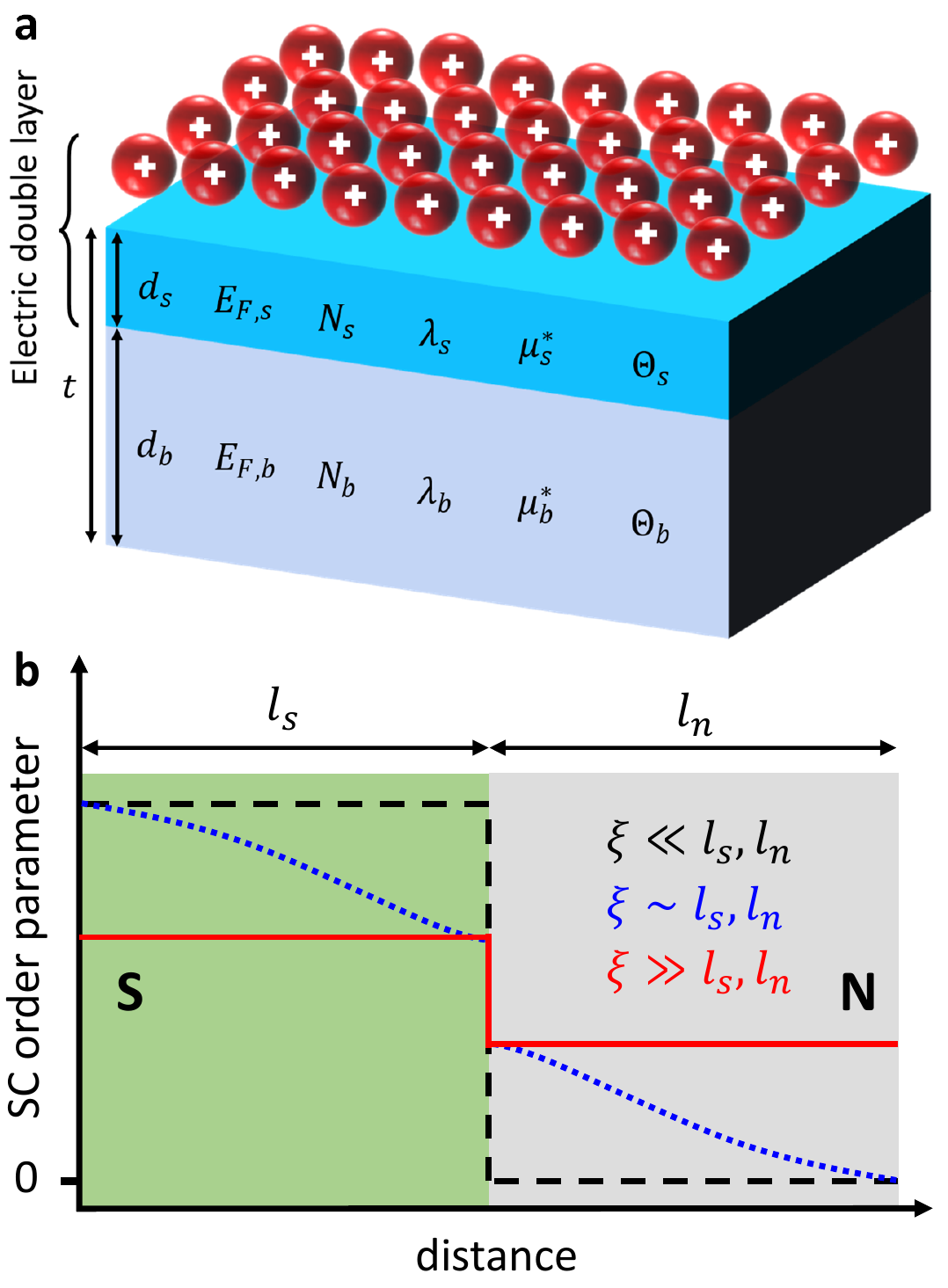}
\caption{
	\textbf{a}, Scheme of an ion-gated superconducting thin film. The EDL is composed by the Helmholtz layer of cations and the charge-doped surface layer (dark blue region). The unperturbed bulk of the film is indicated in light blue color. The input parameters of the proximity-effect model are indicated for both layers (see text for details).
	\textbf{b}, Sketch of the SC order parameter profile (i.e. the SC energy gap) across a superconductor(S)/normal metal(N) junction, based on Ref.~\onlinecite{KlapwijkJS2004}, for three values of the Gor'kov coherence length $\xi$. Dashed black line is the bulk regime ($\xi\ll l\ped{s},l\ped{n}$) where proximity effect is negligible. Dotted blue line is the mesoscopic regime ($\xi \sim l\ped{s},l\ped{n}$) where a sizeable penetration of the superfluid in the normal metal occurs. Solid red line is the Cooper limit ($\xi\gg l\ped{s},l\ped{n}$) described in the text.
}
\label{fig:proximity_effect}
\end{figure}
The apparent contradiction between surface charge doping and bulk modulation of the SC pairing can be solved by considering the coherent nature of the SC condensate. Close to $T\ped{c}$, the superfluid density is small~\cite{HirschPRB2004} and the electrostatic screening is dominated by unpaired electrons~\cite{KoyamaPRB2004}. In the limit of weak and moderate electric fields, the thickness $d\ped{s}$ within which the gate-induced charge doping is confined should then be comparable to the Thomas-Fermi length, i.e. few \AA~at most~\cite{ChazalvielBook}. Conversely, perturbations to the SC order parameter cannot occur over length scales much smaller than the coherence length, which in metallic superconductors is at least of the order of few nanometers and can be as large as few microns~\cite{deGennesRMP1964}. As sketched in Fig.~\ref{fig:proximity_effect}a, gated SC thin films can then be conceptualized from the electronic point of view as a stack of a doped surface layer, where the electronic and vibrational properties are changed by the applied electric field, and an underlying bulk which remains unperturbed. The gate-induced perturbation to the SC pairing can then be described as a coupling between these two layers by means of SC proximity effect~\cite{PiattiPRB2017, UmmarinoPRB2017}.
This is a phenomenon that occurs at any interface between a superconductor and a normal metal, caused by the tunnelling of the Cooper pairs from the superconducting to the normal side (Andreev reflection)~\cite{KlapwijkJS2004}. This leakage of Cooper pairs into the metallic bank can be thought of as an ``evanescent tail" of the SC order parameter, and close to the interface causes the appearance of a finite SC gap in the metal while suppressing it in the superconductor~\cite{deGennesRMP1964, KlapwijkJS2004}, as sketched in Fig.~\ref{fig:proximity_effect}b. If the thicknesses of the two layers are small enough (Cooper limit~\cite{CooperPRL1961}), as is often the case in gated thin films~\cite{ChoiAPL2014, PiattiPRB2017, UmmarinoPRB2017}, the interface can be described as a single superconductor with an effective electron-phonon coupling constant $\langle\lambda\rangle$ that is a weighted average
of the coupling constants in the normal ($\lambda\ped{n}$) and superconducting ($\lambda\ped{s}$) layers~\cite{deGennesRMP1964, KlapwijkJS2004}.

Following the strong-coupling BCS theory for proximity effect~\cite{SilvertPRB1975}, in Ref.~\onlinecite{PiattiPRB2017} it was proposed that the $T\ped{c}$ of a gated thin film could be obtained by:
\begin{equation}
	T\ped{c} = \frac{\Theta}{1.45}\exp\left(-\frac{1+\left\langle \lambda\right\rangle}{\left\langle \lambda\right\rangle -\mu^{*}}\right)
	\label{eq:Tccomb}
\end{equation}
with
\begin{equation}
	\langle\lambda\rangle = \frac{\lambda_{s}N\ped{s}d\ped{s}+\lambda\ped{b}N\ped{b}d\ped{b}}{N\ped{s}d\ped{s}+N\ped{b}d\ped{b}}.
\end{equation}
where the subscripts $s$ and $b$ referred to the surface and bulk layers of the film, $N\ped{s,b}$ are the densities of states (DOS) at the Fermi level, and $d\ped{s,b}$ are the thicknesses of the layers. Of course $d\ped{s}+d\ped{b}=t$ with $t$ the total thickness of the film. It was then assumed that $\Theta$ (the representative temperature of the phonon spectrum) and $\mu^*$ (the Coulomb pseudopotential) were unchanged in the doped surface layer, and that $T_c$ was modulated by charge doping mainly by shifting the Fermi level $E\ped{F}$: This in turn changes the DOS ratio $N_s/N_b$, calculated via density functional theory (DFT), so that $\lambda_{s}=\lambda_{b}\cdot N_{s}/N_{b}$~\cite{PiattiPRB2017}. Despite these rough approximations, the experimentally-measured shifts of the $T\ped{c}$ of gated thin NbN films as a function of both the induced charge density and of the film thickness could be accurately reproduced with the proximity effect model~\cite{PiattiPRB2017}. These findings 
showed that, even when the gate-induced electric field is confined in a thin 
layer at the surface by electrostatic screening, the coherent nature of the SC condensate forces the gate-induced perturbation to the SC state to extend in a region much larger than a single unit cell~\cite{PiattiPRB2017}. Indeed, critical-current measurements in gated mesoscopic samples of metallic superconductors later showed that, at temperatures much lower than $T\ped{c}$, the gate-induced suppressions of the SC state extended even further inside the bulk, approaching the London penetration depth of the material~\cite{DeSimoniNatNano2018, PaolucciPRApp2019, RitterArxiv2020}. In these latter cases, however, it must be noted that the source of the suppression appears to be unrelated with charge doping (which is negligible due to the low capacitance of gating through a vacuum) and is still debated in the literature, with some authors suggesting that the penetration of the electric field may itself be detrimental to mesoscopic superconductors~\cite{DeSimoniNatNano2018, PaolucciPRApp2019}, while others propose instead that the SC state is being destroyed by surface injection of high-energy electrons tunnelling from the gate~\cite{RitterArxiv2020, KavanovArXiv2021}. Critical current enhancements have also been recently reported in gated NbN nanobridges~\cite{RocciNanoLett2021}.

\section{Anomalous electrostatic screening at ultrahigh electric fields}
\begin{figure*}
\centering
\includegraphics[width=\textwidth]{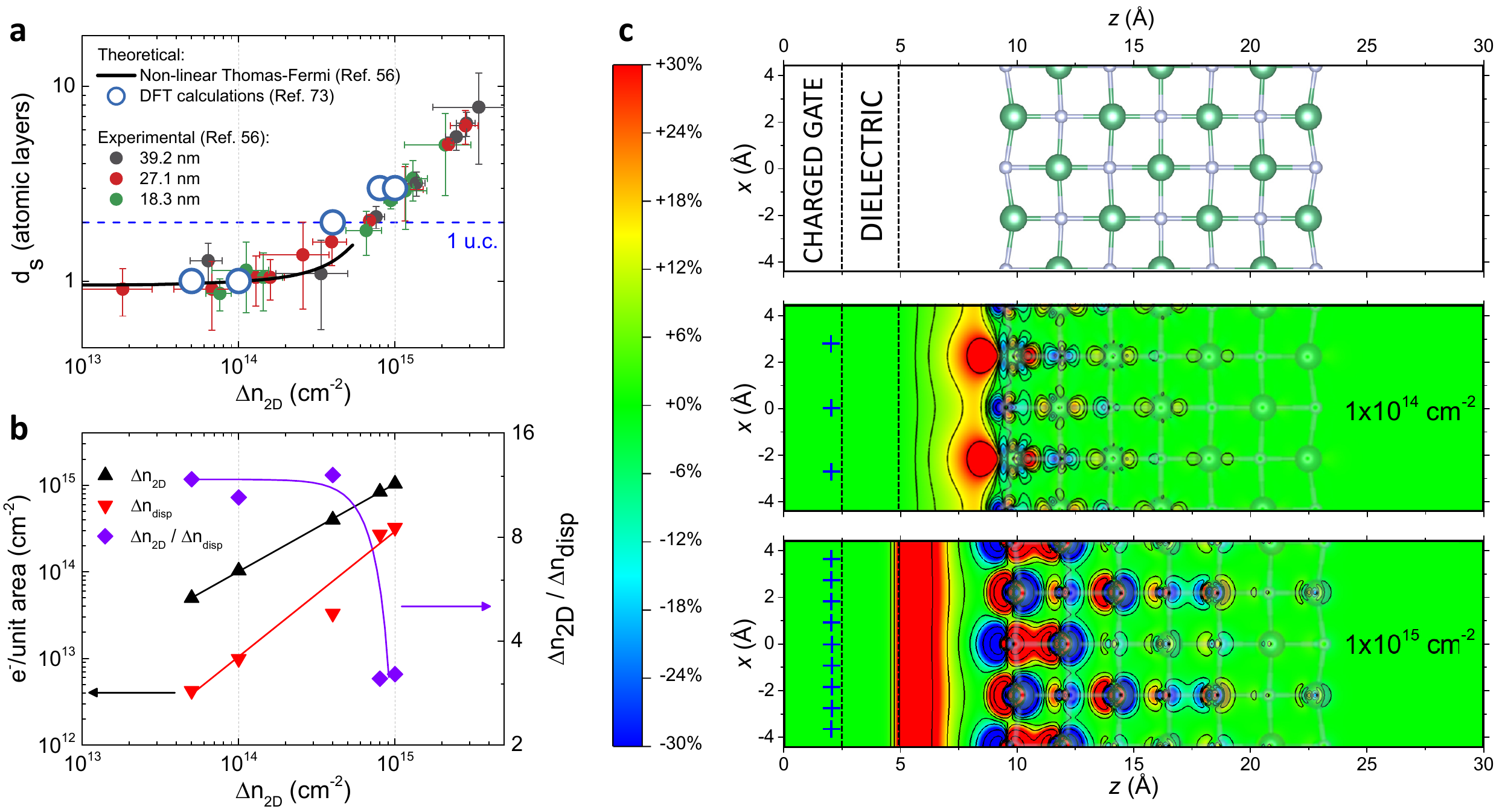}
\caption{
	\textbf{a}, Thickness of the charge-doped surface layer (one atomic layer $\approx 2.2$~\AA) vs. the induced charge density per unit surface in gated NbN thin films. Filled circles are the data from Ref.~\onlinecite{PiattiPRB2017} for different film thicknesses. Solid black line is the result of the non-linear Thomas-Fermi model~\cite{PiattiPRB2017, ChazalvielBook}. Hollow circles are the results from the results from the DFT calculations in Ref.~\onlinecite{PiattiApSuSc2018nbn}. Dashed blue line marks the thickness of one unit cell.
	\textbf{b}, Screening charge contributions per unit surface due to the induced ($\Delta n\ped{2D}$) and displaced ($\Delta n\ped{disp}$) charge densities, together with their ratio $\Delta n\ped{2D}/\Delta n\ped{disp}$, as a function of $\Delta n\ped{2D}$. The solid lines act as guides to the eye.
	\textbf{c}, Ball-and-stick model of the NbN lattice, and color maps of the screening charge in a gated NbN slab for two values of $\Delta n\ped{2D}$ ($1\cdot10^{14}$ and $1\cdot10^{15}$~cm\apex{-2}). The color map is a linear scale between $\pm30\%$ of the maximum value in the surface accumulation layer. \textbf{b},\textbf{c} are reprinted from Ref.~\onlinecite{PiattiApSuSc2018nbn}, with permission from Elsevier.
}
\label{fig:screening}
\end{figure*}

The good agreement between the experimental results on NbN thin films and the strong-coupling BCS model for the proximity effect in Ref.~\onlinecite{PiattiAPL2017} also allowed to map the thickness of the electronically-perturbed surface layer $d\ped{s}$ (i.e. the screening length of the electric field) as a function of the induced charge density and of the film thickness. As shown in Fig.~\ref{fig:screening}a (filled circles), $d\ped{s}$ was found to be independent on the film thickness and smaller than one unit cell at low induced charge density~\cite{PiattiPRB2017}. These two findings are important checks for the consistency of the model, since on the one hand electrostatic screening is a purely surface property that does not depend on the thickness of a material as long as the latter is larger than the screening length~\cite{ChazalvielBook}, and on the other hand a good agreement is found with the standard Thomas-Fermi theory of screening ($d\ped{TF}\sim 2$~\AA~in NbN~\cite{PiattiPRB2017}) which is valid in the limit of low electric fields~\cite{ChazalvielBook}. Surprisingly, however, $d\ped{s}$ was found to strongly increase at large induced charge density (thus, at large gate electric fields), extending for over 5 unit cells for the highest induced values of charge doping~\cite{PiattiPRB2017}, in contrast with the simple Thomas-Fermi picture in which the thickness of a surface accumulation layer is expected to decrease strongly at the increase of the electric field, due to the gate potential well becoming deeper and narrower~\cite{ChazalvielBook, AndoRMP1982}. This was the first time that such an increase in the electrostatic screening length was reported in a conventional metal, although anomalously large thicknesses of the surface accumulation layer had previously been reported in ion-gated YBa\ped{2}Cu\ped{3}O\ped{7}~\cite{FeteAPL2016} and SrTiO\ped{3}~\cite{UenoPRB2014}. Notably, a very similar behavior of $d\ped{s}$ vs. the induced charge density was later reported in SrTiO\ped{3} gated with a standard solid dielectric~\cite{ValentinisPRB2017}, ruling out that such a phenomenon could be associated to ion-gated surfaces only and suggesting that this would be a general behavior of any gated material once sufficiently large electric fields are achieved.

This unexpected feature could be partially accounted for by including non-linear 
corrections to the standard Thomas-Fermi screening model (solid black line in Fig.~\ref{fig:screening}a), but even this extended model breaks down at larger electric fields~\cite{PiattiPRB2017}. A more complete theoretical description was given in Ref.~\onlinecite{PiattiApSuSc2018nbn} by employing a novel DFT method~\cite{BrummePRB2014, BrummePRB2015} that treats the gate-induced charge doping more accurately than a uniform background charge~\cite{GePRB2013, Gonnelli2dMater2017, PiattiApSuSc2020} by fully accounting for the gate electric field of the FET configuration in a self-consistent way~\cite{BrummePRB2016, SohierPRB2017, RomaninApSuSc2019, PiattiJPCM2019, RomaninNC2020, RomaninJAP2020, RomaninApSuSc2021}. The DFT calculations in the FET configuration accurately reproduced the doping-dependence of the thickness of the surface accumulation layer (hollow red circles in Fig.~\ref{fig:screening}a), although they became too computationally costly to investigate the largest values of induced charge density~\cite{PiattiApSuSc2018nbn}. Furthermore, a crucial physical insight on the origin of the breakdown of the Thomas-Fermi approximation could be obtained by considering the spatial dependence of the screening charge for increasing field-induced electron doping. 

As shown in Fig.~\ref{fig:screening}c, the total screening charge was composed of two main contributions, i.e. the accumulation layer where the extra induced carriers were confined, and an additional polarization of the pre-existent electron density~\cite{PiattiApSuSc2018nbn}. The first contribution -- analogous to the Thomas-Fermi screening prediction -- is always localized within the first atomic layer and dominates the total screening charge at low and moderate fields. The second contribution consists instead in the formation of charge dipoles arising from the polarization of the pristine atomic orbitals, and extends for an increasing number of atomic layers from the surface as the gate electric field is increased, eventually reaching well beyond the first unit cell from the surface. This displaced charge density was found to be negligible at low and moderate applied fields, but became comparable to that confined within the accumulation layer upon the application of sufficiently large electric fields~\cite{PiattiApSuSc2018nbn}, as shown in Fig.~\ref{fig:screening}b. 

The findings reported in Ref.~\onlinecite{PiattiApSuSc2018nbn} thus fully confirmed the validity of the proximity-effect model developed in Ref.~\onlinecite{PiattiPRB2017} for how the gate-induced surface perturbation of the electronic properties of a thin film of a metallic superconductor results in a bulk perturbation of its SC properties. Additionally, they also firmly established that a proper treatment beyond the Thomas-Fermi approximation is mandatory to describe the electrostatic screening in the presence of the ultrahigh electric fields typical of the ionic gate technique, even in the case of conventional metallic systems.

\section{Eliashberg theory of gated superconducting films}

\begin{figure*}
\centering
\includegraphics[width=0.8\textwidth]{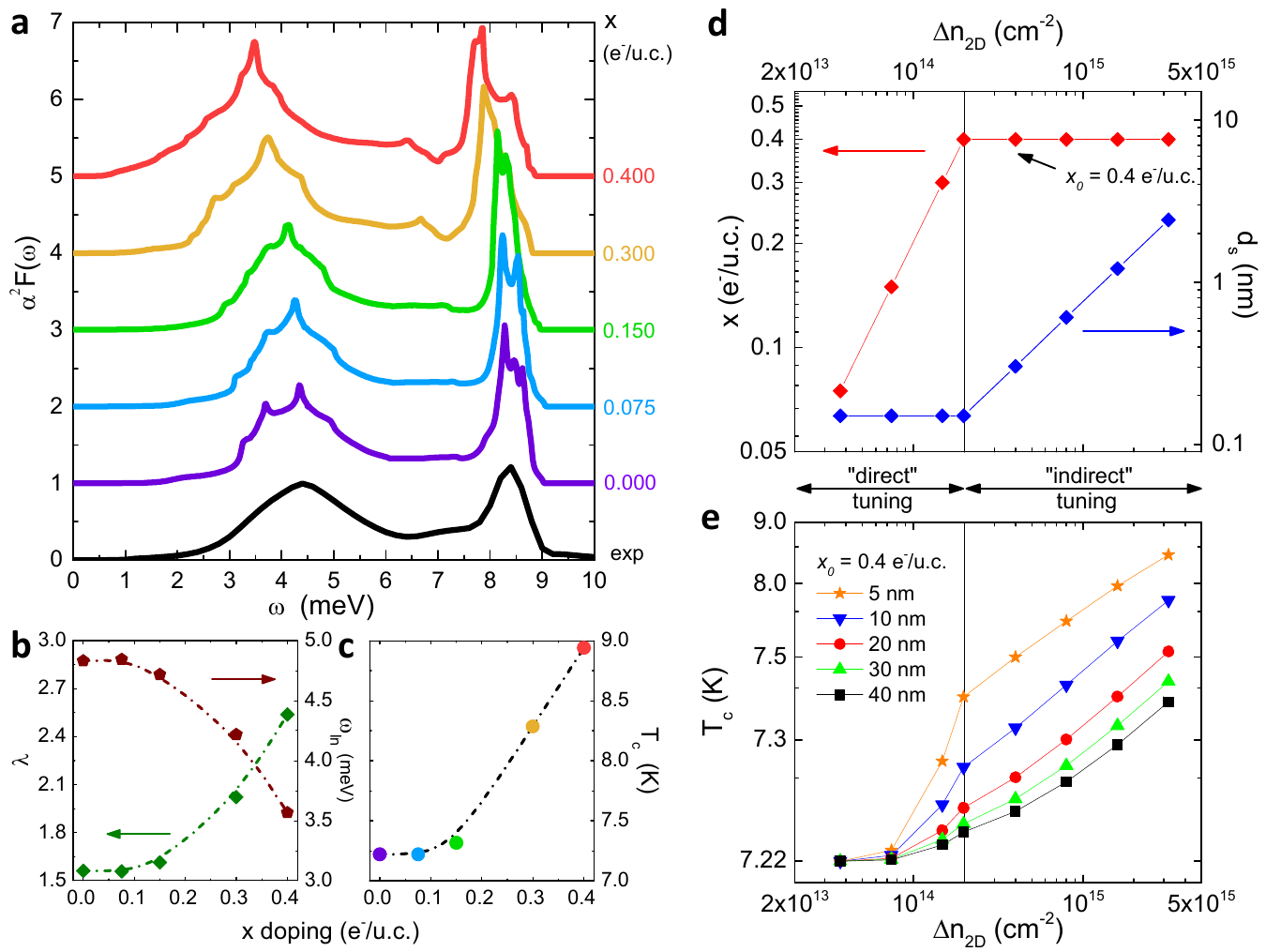}
\caption{
	\textbf{a}, Electron-phonon spectral function vs. frequency for Pb, for different values of charge doping per unit cell $x$. All curves are shifted for clarity by 1. 
	\textbf{b}, Corresponding electron-phonon coupling $\lambda$ (green diamonds) and representative phonon frequency $\omega\ped{ln}$ (brown pentagons) vs. $x$.
	\textbf{c}, Corresponding SC critical temperature $T\ped{c}$ vs. $x$ in the absence of any proximity effect. 
	\textbf{d}, Charge doping per unit cell $x$ and charge-doped surface layer thickness $d\ped{s}$ vs. induced charge density per unit surface used as input for the proximity Eliashberg equations in Ref.~\onlinecite{UmmarinoPRB2017}. 
	\textbf{e}, $T\ped{c}$ vs. induced charge density per unit surface, for different total thicknesses of the gated Pb film. 
	``Direct" and ``indirect" gate-tuning regimes are highlighted in \textbf{d},\textbf{e} (see text for details).
	All lines in \textbf{b}-\textbf{e} are guides to the eye.
	All panels are reprinted with permission from Ref.~\onlinecite{UmmarinoPRB2017}. Copyright 2017 by the American Physical Society.
}
\label{fig:eliashberg}
\end{figure*}

Despite {\color{blue}its} success in describing the observed behavior of gated NbN thin films, the proximity-effect model developed in Ref.~\onlinecite{PiattiPRB2017} described the modulations to the SC state purely in terms of doping-induced changes to the electronic properties of the system, namely the density of states at the Fermi level. While this was proven to be a good approximation for NbN, in general field-effect doping affects both the electronic~\cite{BrummePRB2014, BrummePRB2015} and the vibrational~\cite{SohierPRB2017, RomaninApSuSc2019, RomaninApSuSc2021} properties of a gated material, and a complete model must account for both contributions. This challenge was tackled in Ref.~\onlinecite{UmmarinoPRB2017} by developing a method to determine the $T\ped{c}$ of any gated thin film of a single-band metallic superconductor purely from first principles, by combining DFT calculations with the full Migdal-Eliashberg theory of proximity effect. In this method, the electron-phonon spectral function $\alpha^2F(\omega)$ of the gated material, which describes its frequency-dependent electron-phonon coupling, is determined as a function of charge doping by DFT calculations. The $\alpha^2F(\omega)$ of the doped surface layer and that of the unperturbed bulk are then used as input for the proximity Migdal-Eliashberg equations, which allow to determine $T\ped{c}$ -- and, in principle, also other SC properties -- for the complete structure with arbitrary thickness. If the DFT calculations are performed using a uniform doping~\cite{GePRB2013, Gonnelli2dMater2017, PiattiApSuSc2020, UmmarinoPRB2017}, a selected thickness of the perturbed surface layer must also be included as an input parameter. Conversely, if the DFT calculations are performed with in the FET geometry~\cite{SohierPRB2017, RomaninApSuSc2019, RomaninApSuSc2021}, the model has no free parameters except for the total thickness of the film and the unperturbed value of $\mu^*$. The Eliashberg formalism also allows to directly probe whether the gate-induced tuning of the SC state is mostly caused by a modulation of the electron-phonon coupling $\lambda$ or the representative phonon frequency $\omega\ped{ln}$, since they can both be easily obtained from the $\alpha^2F(\omega)$~\cite{UmmarinoPRB2017}:
\begin{equation}
	\lambda=2\int_{0}^{+\infty}\frac{\alpha^2F(\omega)}{\omega}d\omega
\end{equation}
and
\begin{equation}
	\omega\ped{ln}=\exp\left[\frac{2}{\lambda}\int_{0}^{+\infty}\ln(\omega)\frac{\alpha^2F(\omega)}{\omega}d\omega\right]
\end{equation}
and their doping-dependence compared to that of $T\ped{c}$.

In Ref.~\onlinecite{UmmarinoPRB2017} this method was benchmarked by calculating the $T\ped{c}$ of gated lead (Pb) thin films, as a function of the thicknesses of both the film and the perturbed surface layer. It was first shown that, unlike in In, Nb and NbN thin films, but analogously to Sn thin films, electron accumulation lead to an increase in $T\ped{c}$ (see Fig.~\ref{fig:eliashberg}c). This increase was driven by a strong softening of the phonon modes, as evidenced by the doping-dependence of the $\alpha^2F(\omega)$ (Fig.~\ref{fig:eliashberg}a), which strongly increased $\lambda$ while only weakly reducing $\omega\ped{ln}$ (Fig.~\ref{fig:eliashberg}b)~\cite{UmmarinoPRB2017}. Most importantly, it was found that proximity effect is strongly detrimental to the $T\ped{c}$ tunability achieved by field-effect doping: In the absence of an unperturbed bulk layer, the calculations indicated that the $T\ped{c}$ of Pb could be increased by nearly 2~K for a doping of 0.4 electrons per unit cell. In a 5nm-thick film this increase was already reduced by $\sim90\%$, and by $\sim99.4\%$ in a 40nm-thick film~\cite{UmmarinoPRB2017}.

Using a realistic dependence of the screening length on the induced charge density, mirroring the one found in NbN (see Fig.~\ref{fig:eliashberg}d), it was also shown that the mechanism of $T\ped{c}$ tuning of a superconducting film could be separated in two regimes~\cite{UmmarinoPRB2017}: At low electric fields, where Thomas-Fermi screening holds and the screening length is nearly constant, $T\ped{c}$ is tuned by field-effect doping directly affecting the $\alpha^2F(\omega)$. Whereas at high electric fields, where the Thomas-Fermi model breaks down and the screening length increases with the electric field, $\alpha^2F(\omega)$ remains nearly unchanged and $T\ped{c}$ is tuned via proximity effect only, via the change in the thickness of the perturbed surface layer.
As shown in Fig.~\ref{fig:eliashberg}e, the calculations indicated that this second, indirect tuning allows attaining nearly the same $T\ped{c}$ shift achieved in the absence of the unperturbed bulk layer, but requires nearly 20 times the amount of induced charge density per unit surface~\cite{UmmarinoPRB2017}. In both regimes, increasing the film thickness dramatically reduces the tunability of $T\ped{c}$. As such, the findings reported in Ref.~\onlinecite{UmmarinoPRB2017} clearly indicated how ultrathin films are mandatory to obtain sizeable modulations of $T\ped{c}$ in metallic superconductors, while also reducing the required values of induced charge density -- thus requiring smaller gate voltages to drive the EDL-FETs, in turn allowing to much more easily maintain the electrostatic operation of the ionic gate~\cite{PiattiJSNM2016, PiattiPRB2017}. These results are consistent with the experimental findings reported in gated NbSe\ped{2}, where the $T\ped{c}$ of encapsulated bilayers could be electrostatically tuned by nearly $\approx40\%$ of its original value~\cite{XiPRL2016}.

Beyond the benchmark on gated Pb thin films, which is theoretically well-established but for which no experimental data are yet available, the proximity-Eliashberg method allowed to quantitatively reproduce the original $T\ped{c}$ shift observed in gated indium films~\cite{UmmarinoPSSB2020}, giving a solid theoretical explanation to a problem which had otherwise remained unsolved for nearly fifty years. Furthermore, an extension of this method to two-band metallic superconductor MgB\ped{2} was later developed in Ref.~\onlinecite{UmmarinoJPCM2019}, thus overcoming the strong assumption of the single-band approximation.


\section{Summary and outlook}

In this short review, I have discussed the non-trivial way in which ionic gating can affect the properties of thin films of conventional metallic superconductors. Despite the efficient electrostatic screening of metallic systems confining the gate-induced charge doping to a thin layer at the surface of the material, a bulk-like modulation of the superconducting properties is systematically observed across multiple independent experiments. This coupling of the quasi-2D perturbation of the electronic structure to the long-range ``bulk" modulation of the SC condensate stems from the intrinsically coherent nature of the latter, which cannot be perturbed over length scales smaller than its coherence length. I have examined how this coupling can be described in terms of a proximity effect between the perturbed surface layer and the unperturbed bulk, and discussed how gated thin films of metallic superconductors can be modelled to different degrees of approximation. I have discussed how both experimental findings and ab-initio calculations allowed to unveil an anomalous increase of the electrostatic screening length at the ultra-high electric fields attainable via the ionic gating technique, and reviewed a general framework by which the superconducting properties of any gated thin film of a conventional metallic superconductor can be determined purely from first principles.

A few promising avenues for future investigations may also be suggested. Since the proximity-effect induced ``smearing" of the modulations to the SC pairing is a general phenomenon occurring in any gated superconductor, it could hinder the possibility to obtain large $T\ped{c}$ shifts in any film significantly thicker than the screening length. It is then pivotal to maximize the tunability of the SC state by minimizing the influence of the bulk, such as by employing ultrathin films of metallic superconductors, or by downscaling the devices in more than one direction and investigating gated nanowires, nanobeams or nanoconstrictions, as evidenced by recent experimental findings. It is also necessary to consider device geometries suitable for the measurement of gated SC properties other than the critical temperature, such as the critical current (related to the superfluid density) and the superconducting energy gap. It may also be beneficial to expand the list of metallic superconductors to which ionic gating can be applied, which is often hampered by susceptibility to irreversible electrochemical interactions with the electrolyte: In this context, the development of a scalable encapsulation technique that would preserve a high gate capacitance would be needed. 

Finally, in the perspective of potential applications the capability to operate the gate at temperatures close or lower than the critical temperature of the superconducting film would be highly desirable, demanding further investigations towards the development of electrolytes with lower melting point, or highly reliable ferroelectric gates. {\color{blue}This is because the necessity to warm up the devices above their transition temperature to change the ionic-gate polarization limits applications to reconfigurable SC circuits and devices such as resonators~\cite{resonators}. However, attaining sizeable gate modulations below their transition temperature would open up a plethora of tantalizing applications, including low-dissipation logic and memory elements~\cite{RocciNanoLett2021}, novel architectures for qubits, interferometers, amplifiers and photodetectors~\cite{PaolucciPRApp2019}, as well as magnetometers and heat-mastering systems~\cite{PaolucciAVSQS2019}.}

\end{document}